# High Velocity Dispersion in a Rare Grand Design Spiral Galaxy at Redshift z =2.18

David R. Law[1], Alice E. Shapley[2], Charles C. Steidel[3], Naveen A. Reddy[4], Charlotte R. Christensen[5], Dawn K. Erb[6]

[1] Dunlap Institute for Astronomy & Astrophysics, University of Toronto, 50 St. George Street, Toronto M5S 3H4, Ontario, Canada

[2] Department of Physics and Astronomy, University of California, Los Angeles, CA 90095

[3] California Institute of Technology, MS 249-17, Pasadena, CA 91125

[4] Department of Physics and Astronomy, University of California, Riverside, CA 92521

[5] Steward Observatory, 933 North Cherry Ave, Tucson, AZ 85721

[6] Department of Physics, University of Wisconsin-Milwaukee, Milwaukee, WI 53211

**Although relatively common in the local Universe, only one grand-design spiral galaxy has been spectroscopically confirmed[1] to lie at z>2 (HDFX 28; z=2.011), and may prove to be a major merger that simply resembles a spiral in projection. The rarity of spirals has been explained as a result of disks being dynamically 'hot' at z>2 [2,3,4,5] which may instead favor the**

**formation of commonly-observed clumpy structures [6,7,8,9,10]. Alternatively, current instrumentation may simply not be sensitive enough to detect spiral structures comparable to those in the modern Universe[11]. At redshifts <2, the velocity dispersion of disks decreases[12], and spiral galaxies are more numerous by z~1 [7,13,14,15]. Here we report observations of the grand design spiral galaxy Q2343-BX442 at z=2.18. Spectroscopy of ionized gas shows that the disk is dynamically hot, implying an uncertain origin for the spiral structure. The kinematics of the galaxy are consistent with a thick disk undergoing a minor merger, which can drive the formation of short-lived spiral structure [16,17,18]. A duty cycle of < 100 Myr for such tidally-induced spiral structure in a hot massive disk is consistent with their rarity.**

Using Hubble Space Telescope (*HST*) Wide-Field Camera 3 (WFC3) infrared imaging data tracing rest-frame ~ 5000 Å stellar continuum emission (see details in the Supplementary Information) we find that Q2343-BX442 (hereafter BX442) is well resolved with a total luminous radius R ~ 8 kpc, prominent spiral arms, a central nucleus, and a faint companion located 11 kpc distant in projection to the northeast. These morphological characteristics (see summary in Table 1) lead us to tentatively identify BX442 as a late-type Sc grand design spiral galaxy. Strikingly, BX442 is the only object to display regular spiral morphology out of a sample of 306 galaxies with similar imaging[10] at roughly the same redshift. We used the Keck/OSIRIS spectrograph in concert with the laser guide star adaptive optics (LGSAO) system to obtain integral field spectroscopy of nebular Hα emission from ionized gas regions within BX442 at an angular resolution comparable to that of the *HST* imaging data (~ 2 kpc; see details in Supplementary Information). As illustrated in Figure 1, the ionized gas emission similarly traces the structure of the spiral arms but does not exhibit a prominent central bulge.

Stellar population model fits to broadband photometry (see details in Supplementary Information) indicate that the total star-formation rate (SFR) of BX442 is $52^{+37}_{-21}$ $M_\odot$ yr$^{-1}$, with a characteristic population age of $1100^{+1000}_{-500}$ Myr and visual extinction E(B−V) = 0.3 ± 0.06. BX442 is therefore drawn from the high end of the stellar mass function of z ~ 2 star-forming galaxies, with correspondingly higher than average size, dust extinction, stellar population age, and SFR. This star formation is concentrated in the spiral arms, suggesting that their stellar populations may be significantly younger than those of the nucleus and inter-arm regions. Extrapolating from the Hα line flux map using the Kennicutt relation[19,20], we estimate that the mean SFR surface density in the arms is $\Sigma_{SFR}$ = 0.4 $M_\odot$ yr$^{-1}$ kpc$^{-2}$, peaking at ≈ 1 $M_\odot$ yr$^{-1}$ kpc$^{-2}$ in a bright clump located in the northern arm. Although these values are only modest in comparison to characteristic values of $\Sigma_{SFR}$ ~ 1 -10 $M_\odot$ yr$^{-1}$ kpc$^{-2}$ that have been observed in detailed studies of the star-forming clumps of z ~ 2 galaxies[21], they are nonetheless ~ 30 times greater than typical for local spiral galaxies and are similar to the values observed in local circumnuclear starbursts[19]. As indicated by Keck/LRIS rest-UV spectroscopy (see details in Supplementary Information), the high $\Sigma_{SFR}$ of BX442 drives outflows of gas into the surrounding intergalactic medium with speeds up to a few hundred km s$^{-1}$, although there is no indication that the highest $\Sigma_{SFR}$ regions are the specific launching sites for galactic-scale outflows (see Supplementary Figure S5) as recently proposed for a similar sample of galaxies[21].

Fitting a Gaussian profile to the Hα emission line at each location across the galaxy we determined that the velocity profile of BX442 (Figure 2) is consistent with the rotating disk hypothesis, and exhibits a smooth gradient of ±150 km s$^{-1}$ along the morphological major axis with flux-weighted mean velocity dispersion $\sigma_m$ = 66 ± 6 km s$^{-1}$ (after correcting for the instrumental resolution). The faint companion detected in the *HST* image is spectroscopically confirmed to lie within 100 km s$^{-1}$ of the systemic redshift of BX442 but does not follow the global rotational velocity field, and may therefore represent a merging dwarf galaxy with a mass a few percent that of the primary (as determined from the rest-frame ~ 5000 Å luminosity ratio). The velocity dispersion of the ionized gas in BX442 is highest in the spiral arms and appears to peak at $\sigma$ = 113 ± 14 km s$^{-1}$ in a bright star-forming clump in the northern arm.

We constructed a three-dimensional inclined disk model (see details in Supplementary Information) that accounts for observational effects such as the delivered point spread function, and determined that BX442 is consistent (reduced chi-square $\chi^2_r$ = 2.3) with being a rotating disk inclined at 42° ± 10° to the line of sight with an inclination-corrected circular velocity $v_c$ =234$^{+49}_{-29}$ km s$^{-1}$ at the outer edge of the disk (R ~ 8 kpc). As inferred from our best-fit model, the vertical velocity dispersion of the disk is $\sigma_z$ = 71 km s$^{-1}$ ($v_c/\sigma_z$ ≈ 3), indicating that the system is geometrically thick with a scaleheight $h_z = \sigma_z^2/(2\pi G\Sigma)$ = 0.7 kpc that is comparable to similarly massive systems studied in the literature[2,22,23]. The implied dynamical mass of BX442 is $M_{dyn}$ = 1.0$^{+0.5}_{-0.2}$ x 10$^{11}$ M$_\odot$ within a radius of R = 8 kpc, consistent with the sum of estimates of the gas

($M_{gas} = 2^{+2}_{-1} \times 10^{10}$ $M_\odot$) and stellar ($M_* = 6^{+2}_{-1} \times 10^{10}$ $M_\odot$) masses estimated from inversion of the Schmidt-Kennicutt law[19,20] and stellar population modeling respectively.

Contrary to expectations[9,11,13], BX442 indicates that dynamically hot z ~ 2 disk galaxies can both form spiral structure and that such structure can be easily detected with current-generation instruments such as *HST*/WFC3. Indeed, despite its high velocity dispersion, the mass surface density of BX442 is sufficiently high that the Toomre parameter[24] $Q \leq 1$ throughout the majority of the disk (see details in Supplementary Information), suggesting that BX442 is susceptible to spontaneous formation of spiral structure. Galaxies with physical properties similar to BX442 are not remarkably uncommon at z~2; large samples with similar physical characteristics have been studied using high angular resolution imaging[25,26], integral-field spectroscopy[2,4,5], or both[5,27]. In particular, 27 galaxies in the recent morphology survey from which BX442 was drawn[10] have stellar masses within a factor of 2 of its mass, 10 of which also have similar half-light radii, star-formation rates, dust content, and stellar population ages. None of these other systems have clear spiral structure, indicating either that the triggering mechanism is relatively rare or that the duty cycle of the spiral pattern is extremely short.

Perhaps the most obvious distinction of BX442 is the fact that it appears to be currently experiencing a close-passage minor merger, which numerical simulations and theoretical calculations suggest can be a natural means of producing grand design spiral patterns in galactic disks[16,18,24], even for mass ratios as modest as a few percent[17]. Indeed, many of the most well-known grand design spiral galaxies in the nearby universe (e.g., M51, M81, M101) are observed to have nearby companions, and small satellites such as the Sagittarius (Sgr) dwarf galaxy may even be partly responsible for producing spiral patterns in our own Milky Way galaxy[28]. We test the plausibility of the merger-induced scenario by comparing BX442 to a z ∼ 2 model galaxy selected from a set of extremely high-resolution *N*-body smoothed particle hydrodynamic simulations[29] (see details in Supplementary Information). Although the model disk spontaneously forms flocculent spiral structure in isolation, the lifetime of grand-design spiral patterns induced by the merging companion is generally less than half a rotation period (i.e., ≤ 100 Myr, or $\Delta z \leq 0.08$ for BX442).

Such a mechanism may therefore naturally explain why visible spiral structure at z ∼ 2 is so rare: Not only must a galaxy be sufficiently massive to have stabilized the formation of an extended disk[30], but this disk must then be perturbed by a merging satellite sufficiently massive and properly oriented as to excite an observable grand-design spiral pattern. Further, this spiral must be observed in the narrow window of time for which its strength is at a maximum, and must be oriented sufficiently close to face-on that the pattern is recognizable.

Supplementary Information is linked to the online version of the paper at www.nature.com/nature.

Acknowledgements: DRL and CCS have been supported by grant GO-11694 from the Space Telescope Science Institute, which is operated by the Association of Universities for Research in Astronomy, Inc., for NASA, under contract NAS5-26555. AES acknowledges support from the David and Lucile Packard Foundation. CRC acknowledges support from the US National Science Foundation through grant AST-1009452. DRL appreciates discussions with J. Taylor, R. Abraham, J. Dubinski, F. Governato, and A. Brooks, and thanks M. Peeples for assistance obtaining the Keck/OSIRIS data.

Author Contributions: D.R.L. performed the morphological analysis of the Hubble Space Telescope data and wrote the main manuscript text. The Keck/OSIRIS data was obtained by D.R.L. and A.E.S. and analyzed by D.R.L. with extensive input from A.E.S. and C.C.S.. N.A.R. provided the Keck/LRIS spectra, Spitzer/MIPS photometry, and stellar population modeling code, C.R.C. contributed the hydrodynamic galaxy simulations, and D.K.E. provided the Keck/NIRSPEC spectra. All authors reviewed, discussed, and commented on the manuscript.

Author Information: Correspondence should be addressed to D.R.L. (drlaw@di.utoronto.ca)

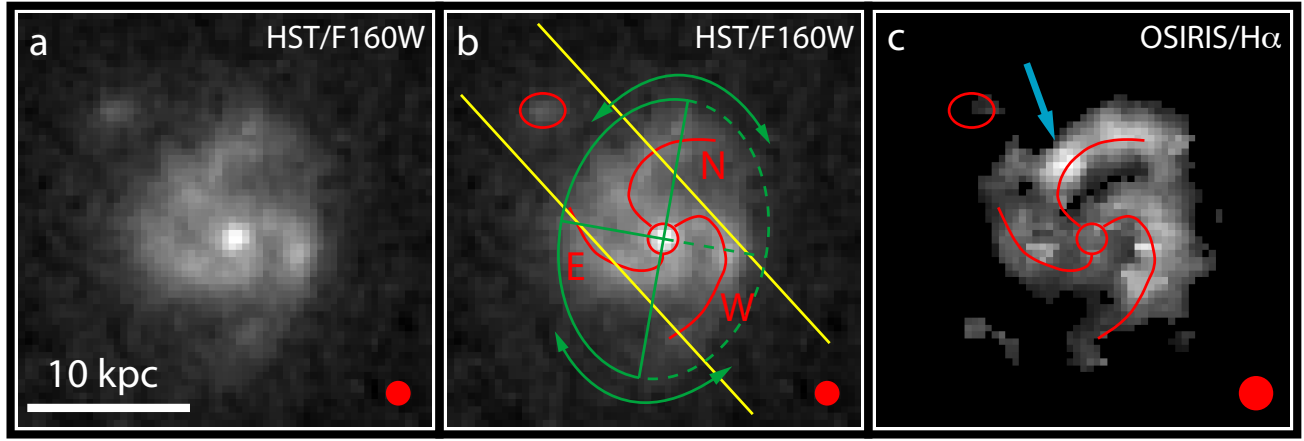

**Fig. 1. Broadband and spectral emission line morphology of BX442.** Panels a and b show the HST/WFC3 F160W broadband morphology; overlaid on panel b are red lines denoting the locations of the northern (N), western (W), and eastern (E) spiral arms, core, and nearby satellite companion, green lines indicating the orientation of the best-fit inclined disk model (solid/dashed green lines represent opposite sides from the midplane), and yellow lines representing the orientation of the long slit for previous Keck/NIRSPEC spectroscopy. The locations of these overlaid lines are defined visually and are intended simply to guide the eye; the alignment of individual images is discussed in §1.4 of the Supplement. Panel c shows the Keck/OSIRIS Hα emission line flux map overlaid with the red lines from panel b, along with a blue arrow indicating the location of a bright star-forming clump in the northern arm. A visual rejection criterion roughly correspondingly to a cut of signal-to-noise ratio > 3 (see details in Supplementary Information) was used to mask low-flux pixels. Each panel is 3 arcsec to a side, corresponding to 25.3 kpc at the redshift of BX442, oriented with North up and East to the left. The red dot in all panels indicates the FWHM of the observational point spread function.

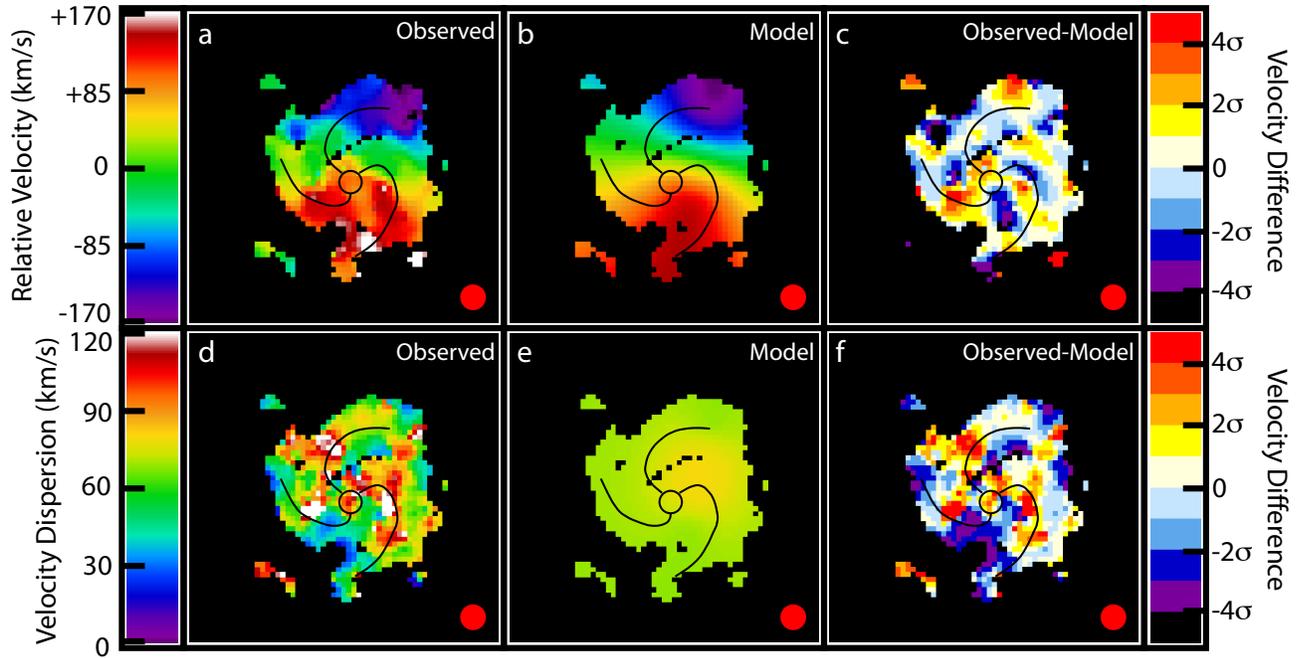

**Fig 2. Kinematic velocity and velocity dispersion maps of BX442.** Panels a and d show the observed relative velocity and velocity dispersion maps (uncorrected for instrumental resolution) recovered from fitting Gaussian emission line profiles to the Hα emission in each spatial pixel. The total integration time was 13 hours with a point-spread function (PSF) width measuring 0.25 arcsec (corresponding to ∼ 2 kpc at the redshift of BX442). Panels b and e respectively show the best-fit inclined-disk models of the relative velocity and velocity dispersion (after convolution with the observational PSF and R ∼ 3100 spectral resolution), while panels c and f show the residuals of the observed velocity and velocity dispersion fields minus the best-fit model. The residual values are given in units of the observational uncertainty; 1σ corresponds to 17 km s$^{-1}$ for the line-of-sight velocity, and 14 km s$^{-1}$ for the line-of-sight velocity dispersion. Black lines indicating the spiral disk structure are overlaid from panel b of Figure1; these lines indicate that the kinematic center of BX442 is offset from the apparent nucleus of the continuum flux by ∼ 2 kpc, due in part to the uncertainty in image registration between the HST/WFC3 and Keck/OSIRIS data (see discussion in Supplementary Information). The red circles indicate the FWHM of the observational point spread function.

**Table 1: Physical Characteristics of BX442.**

| | | | |
|---|---|---|---|
| R.A. (J2000) | 23:46:19.35 | Bulge/Total ¶ | 10% |
| Decl. (J2000) | +12:48:00.0 | Inclination | 42° ± 10° |
| Redshift | 2.1765±0.0001 † | Pitch Angle § | 37° ±6° |
| Lookback time | 10.7 Gyr | Spiral Arm Contrast ¥ | 1 AB arcsec$^{-2}$ |
| Stellar Mass | $6^{+2}_{-1} \times 10^{10}$ M$_\odot$ | Circular velocity | $234^{+49}_{-29}$ km s−1 |
| Gas Mass | $2 \times 10^{10}$ M$_\odot$ | Optical Radius | 8 kpc |
| Age | $1100^{+1000}_{-500}$ Myr | Velocity Dispersion | 71±1 km s$^{-1}$ |
| SFR$_{SED}$ | $52^{+37}_{-21}$ M$_\odot$ yr$^{-1}$ | Hubble Type | Sc |
| SFR$_{H\alpha}$ | 45 M$_\odot$ yr$^{-1}$ | | |

See Supplementary Information for details.

† Derived from Hα nebular line emission and confirmed by multiple other emission and absorption line features.

§ Fourier phase profile analysis of the spiral arms indicates substantial power in the m = 2 and m = 3 symmetry modes, corresponding to a three-armed spiral pattern (in which one arm is foreshortened by the inclination to the line of sight) with opening pitch angle α = 37° ±6°

¶ Decomposing the central nucleus from the surrounding disk using a model of the *HST*/WFC3 point spread function indicates that the nuclear emission region contributes ∼10% of the total rest-frame ∼ 5000 Å continuum flux, and has Sersic radial profile index n ≥ 4 and a half-light radius r ≤ 1.5 kpc consistent with galactic bulges in nearby disk galaxies.

¥ Spiral arm/interarm surface brightness differential.

**Supplementary Information**

In this supplementary information we provide additional details regarding the data collection, analysis, and interpretation of BX442. In §1, 2, 3, and 4 we describe the observational characteristics of the *HST*/WFC3 imaging data, Keck/LRIS rest-UV spectroscopy, Keck/NIRSPEC long-slit rest-optical spectroscopy, and Keck/OSIRIS rest-optical IFU spectroscopy respectively. We discuss our methods for fitting stellar population models and determining gas masses in §5, comparing these global estimates with spatially resolved maps of stellar population age determined from the ratio of the emission line and continuum flux maps. Our approach to developing the best-fit rotating disk model is described in §6, and we discuss detailed calculations of the spatially variable [N II]/H$\alpha$ nebular emission line ratio in §7. Finally, we detail our analysis of the dynamical stability in terms of the Toomre $Q$ parameter in §8 and discuss corresponding smoothed-particle hydrodynamic (SPH) simulations in §9. Throughout our calculations we adopt a standard $\Lambda$CDM cosmology based on the seven-year WMAP results[31] in which $H_0 = 70.4$ km s$^{-1}$ Mpc$^{-1}$, $\Omega_M = 0.272$, and $\Omega_\Lambda = 0.728$.

1. *HST* IMAGING DATA

BX442 was observed on June 14, 2010 with the WFC3 camera on board *HST* as part of the survey program GO-11694 (PI: Law). The goals and observational details of this program are described in detail elsewhere[10]; in brief, the survey was designed to obtain rest-frame optical morphological data for a sample of 306 spectroscopically-confirmed star-forming galaxies in the redshift range $z = 1.5 - 3.6$ along the line of sight to bright background QSOs. BX442 was originally identified as a $z \sim 2$ candidate by its rest-UV color[32–34] and spectroscopically confirmed to lie at $z \sim 2.18$ using Keck/LRIS rest-UV and Keck/NIRSPEC rest-optical spectroscopy.

The *HST*/WFC3 data for Q2343-BX442 consisted of 3 orbits of integration using the F160W filter ($\lambda_{\text{eff}} = 15369$ Å, which traces rest-frame $\sim 5000$ Å at $z = 2.18$) for a total integration time of 8100 seconds composed of nine 900 second exposures dithered using a nine-point sub-pixel offset pattern designed to uniformly sample the PSF. The data were reduced using the MultiDrizzle[35] software package to clean, sky subtract, distortion correct, and combine the individual frames. The raw WFC3 frames are undersampled with a pixel scale of 0.128 arcsec; in order to aid comparison to the Keck/OSIRIS data these frames were drizzled to a pixel scale of 0.05 arcsec pixel$^{-1}$ using a pixel droplet fraction of 0.7. The point spread function of the WFC3 data is found to be $0.18 \pm 0.01$ arcsec based on isolated and unsaturated stars in the imaging fields. Details regarding photometric calibration are given elsewhere[10]; the data have a $3\sigma$ surface brightness sensitivity of 25.1 AB arcsec$^{-2}$.

1.1. *Morphological Fourier Decomposition*

It is difficult to determine visually whether BX442 is a two-component or three-component spiral, and we therefore quantify the degree of spiral structure using a Fourier analysis technique[36,37] to decompose the galaxy into a superposition of $m$-armed logarithmic spirals. First, we deprojected the *HST* image of BX442 to how it would appear face-on using the position angle and inclination derived in §6 . We then calculate the Fourier amplitudes

$$A(p,m) = \frac{1}{D} \sum_{j=1}^{N} f_j e^{-i(pu_j + m\theta_j)} \quad (1)$$

where $D = \sum_{j=1}^{N} f_j$ is the sum of all $N$ pixel fluxes $f_j$ in the deprojected image, $u_j$ is the logarithm of the pixel radius from the galaxy center (defined by the central nucleus), $\theta_j$ is the angular position of a given pixel $j$, and $p$ is related to the spiral pitch angle $\alpha$ by $p = -m/\tan(\alpha)$. In Figure S1 we plot the Fourier amplitude as a function of $p$ for $m = 1 - 4$ and radii in the range $r = 2 - 8$ kpc (i.e., excluding the central nucleus and comparable to the $r_{1/2} \sim 5$ kpc half-light radius of the disk).

There is significant power in the $m = 1$, 2, and 3 Fourier modes, indicating the complex structure of BX442 and the non-uniform surface brightness of its spiral arms. As indicated by Figure S1, the majority of the power is in the $m = 2$ mode, which peaks at $p = 0$ corresponding to an approximately linear (i.e., bar-like) distribution of flux in the deprojected image. This arises due to the close proximity between the northern (N) and eastern (E) arms and the faintness of the outer regions of the western (W) arm (see Figure 1), resulting in a predominantly linear distribution of flux along the southeast-northwest axis. Similarly, the faintness of the W arm compared to the N arm gives rise to a measurable $m = 1$ (i.e., lopsided) distribution of the light profile. The outer regions of the spiral arms where the curvature can be seen more clearly is apparent in the $m = 3$ solution, whose peak amplitude occurs at $p = 4$, corresponding to a spiral arm pitch angle (i.e., the 'opening angle') of $\alpha = 37°$.

We evaluate the impact of the data quality on the derived Fourier components by performing numerical simulations accounting for the angular resolution and pixel noise characteristics of our data. In brief, we used GALFIT[38,39] to create a model of a three-armed spiral galaxy that has a central bulge and spiral arms that approximately match BX442 in terms of their winding, radial extent, and arm/interarm surface brightness ratio when convolved with the *HST*/WFC3 point spread function. This model galaxy was scaled to the total F160W magnitude of BX442 ($H_{160} = 22.04$ AB) and placed in a blank region of the WFC3 imaging field to introduce realistic noise. The Fourier profile of the noisy, PSF-smeared model was not significantly different from the profile of the base model (i.e., with zero noise and angular resolution limited only by the pixel scale); although the introduction of a realistic PSF and pixel noise produced some variation in the power spectra, there was no significant aliasing of power between $m$ modes, and the intrinsic pitch angle of the spiral was recovered to an accuracy of $\pm 6°$.

We therefore conclude confidently that BX442 has morphology broadly characteristic of a three-armed spi-ral with $\alpha = 37° \pm 6°$, and that additional features in the Fourier spectra genuinely represent complexities in its detailed structure rather than simply uncertainties inherent to our *HST*/WFC3 imaging data. Such three-armed spirals are rare in the nearby universe although some genuine examples have been noted (e.g., NGC 7309, NGC 2997)[40,41] and even traditional grand-design spirals such as M 51 may contain $m = 3$ mode structures[42]. The three-armed structure may occur as a result of either self-generated instabilities or possibly interaction with a nearby companion (since the number and morphology of the arms will depend on the detailed properties of the merger), and may also indicate a physical difference in the stability modes of spiral structure in the young universe where galactic disks have significant vertical velocity dispersions and correspondingly different epicyclic harmonics.

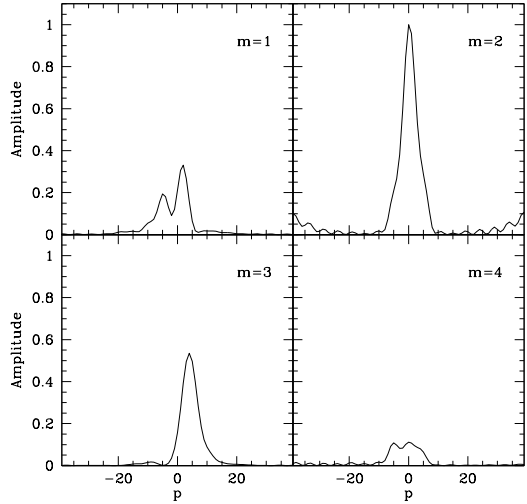

FIG. S1.— **Normalized amplitudes of the Fourier spectra coefficients $A(p,m)$ for BX442.** Fourier coefficients are calculated over the range of radii $r = 2 - 8$ kpc. There is significant power in the $m = 2$ and $m = 3$ modes, corresponding to spiral structure with three-fold rotational symmetry.

## 2. KECK/LRIS REST-UV SPECTROSCOPY

BX442 was observed using the Keck/LRIS spectrograph on 08 October 2010 for a total integration time of $\sim 7$ hours in good conditions. The spectrum was reduced, flux-calibrated[32], and shifted to the systemic rest frame using the OSIRIS H$\alpha$ nebular emission line redshift $z_{H\alpha} = 2.1765 \pm 0.0001$. As illustrated in Figure S2, BX442 has negligible Ly$\alpha$ emission and multiple deep absorption features reaching zero relative flux (implying that most of the starburst emitting regions are covered by cool interstellar gas). The absorption features exhibit extended blue wings characteristic of a strongly outflowing wind reaching up to a few hundred of km s$^{-1}$, similar to those observed in other star-forming galaxies at $z \sim 2$[43]. The mean blueshift of the absorption lines $\Delta v_{IS} = -38$ km s$^{-1}$ however, indicating that a substantial portion of the absorbing gas in BX442 lies at or near the systemic redshift, consistent with expectations for an extended star-forming disk[43] (D.R.L. et al. 2012, manuscript in preparation).

It is probable that the outflowing wind of BX442 has driven substantial quantities of gas to beyond the virial radius. As discussed previously in the literature[44], Ly$\alpha$ and C IV absorption features at the redshift of BX442 are observed in the rest-UV spectrum of the background galaxy Q2343-BX418 ($z = 2.3$) with an impact parameter of 17 arcsec (corresponding to 140 kpc at the redshift of BX442). The C IV absorption feature is stronger (equivalent width 2Å) than average for the parent $z \sim 2$ star-forming galaxy population[43], consistent with the high mass and old age of BX442 (§5).

## 3. KECK/NIRSPEC REST-OPTICAL LONG-SLIT SPECTROSCOPY

BX442 was observed in the $J$, $H$, and $K$ bands with the Keck/NIRSPEC long-slit spectrograph between June and September 2004 as part of an observational campaign targeting the nearby galaxy Q2343-BX418. The details of these observations are presented elsewhere[44]. As illustrated by Figure S3, H$\alpha$, [N II], [O III], and [O II] emission from BX442 are all well detected, although only an upper limit is obtained on the H$\beta$ emission line flux (possibly because it falls near a bright OH atmospheric emission line).[7] The total emission line fluxes and corresponding dust-corrected luminosities for each of these lines are tabulated in Table S1. These spectra did not reveal strong evidence for velocity shear across BX442[45] because the NIRSPEC slit was misaligned with the kinematic major axis of the galaxy by $\sim 50°$. Indeed, BX442 was only observed due to its angular proximity to another galaxy, Q2343-BX418, and in the absence of spatially resolved information the position angle of the NIRSPEC slit ($\phi_{\rm NIRSPEC} = 43°$ East of North) was chosen to include both galaxies.

## 4. KECK/OSIRIS REST-OPTICAL IFU SPECTROSCOPY

BX442 was observed on the nights of August 23-25 2011 using the OSIRIS integral-field spectrograph[46] in combination with the Keck II laser guide star adaptive optics (LGSAO) system. We used the 100 mas lenslet scale in combination with the Kn2 narrowband filter (resulting in a FOV $\sim 4 \times 6$ arcsec) in order to observe H$\alpha$ emission redshifted to $\lambda = 20852.4$ Å. Our observing strategy has already been described at length in the literature[5,3]; in brief, the target was acquired by means of a blind offset from our tip-tilt reference star ($R \sim 16.2$, at a separation of 46.1 arcsec and position angle of 180°). Observations were made in AB pairs of 900 seconds per frame where the galaxy was alternately placed on the top or bottom of the IFU, with small ($\sim 1-2$ lenslet) random dithers about each location. A total of 52 frames were obtained in such a manner over three nights of observing, resulting in a total of 13 hours on-source integration time. Conditions were photometric for all three nights, with median $V$-band seeing $\sim 0.4 - 0.5$ arcsec. Standard star observations of HIP 111538 (HIP 16652) were taken at the beginning (end) of each night, and the tip-tilt guide star was observed during each acquisition to confirm proper image centering and to assess the AO-corrected PSF.

---
[7] It is not possible to estimate a limit to the nebular extinction $E(B-V)_{\rm neb}$ using the H$\alpha$/H$\beta$ line ratio due to systematic uncertainties in the slit losses of the $H$ and $K$ band spectra that were obtained months apart in different seeing conditions.

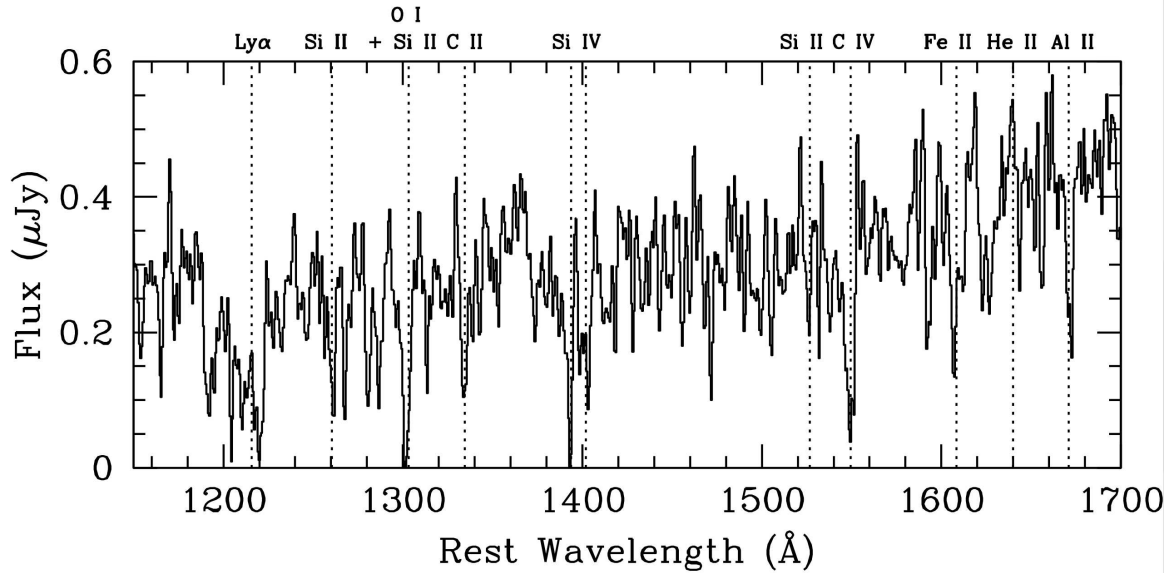

Fig. S2.— **Flux-calibrated rest-UV spectrum of BX442.** Vertical dotted lines indicate the fiducial wavelengths of major atomic transitions.

Raw observational data were reduced and combined using custom IDL routines and following procedures previously described in the literature[5]. We first combined five individual 900-second dark frames to construct a super-dark reference frame using the OSIRIS pipeline routine "COMBINE FRAMES." This super-dark frame was subtracted from each science frame, and the resulting frames were extracted into data cubes using the standard OSIRIS pipeline algorithm, which performs channel level adjustment, glitch and cosmic ray identification, and removal of crosstalk. The A and B frames in each AB pair of observations were then differenced from each other, scaling the subtracted frame to match the median of the object frame in each spectral channel. Each of the resulting frames was flux and telluric calibrated using standard star observations taken closest to it in time and airmass. Each of these 52 cubes were stacked together using a $3\sigma$ clipped mean algorithm. Spaxels (spatial elements derived from individual lenslets) in the stacked image with contributions from fewer than 75% of the input frames were masked from the final image in order to ensure that no negative image residuals are present in the stacked data cube. This stacked data cube was spatially resampled to 50 mas/pixel and lightly smoothed with a Gaussian kernel of FWHM 0.16 arcsec at each spectral channel in order to increase the quality of the spectra in each spaxel. The final size of the reduced data cube is $60 \times 60$ pixels, corresponding to $3 \times 3$ arcsec.

Reducing our observations of the tip-tilt reference star in a manner analogous to the BX442 science frames, we determine from the combined, oversampled, and smoothed tip-tilt star cube that our final PSF has FWHM 0.25 arcsecond.[8] Using isolated OH emission lines in the science frames we estimate that the average[9] spectral resolution of the observations is $R \sim 3100$.

Efforts to extract kinematic and flux information from the reduced data cube using automated routines gave suboptimal results, frequently confusing faint and/or narrow emission line features with noise spikes corresponding to OH atmospheric emission line residuals. We therefore inspected each of the 3600 spectra in the reduced data cube by hand, and constructed flux, velocity, and velocity dispersion maps by fitting the H$\alpha$, [N II] $\lambda 6549$, and [N II] $\lambda 6585$ emission features interactively with single Gaussian profiles using the IRAF task *splot*. Such single Gaussian profiles were adequate models for the observed emission line profile at each location throughout the galaxy (even in the vicinity of bright star-forming clumps), with no evidence for marked asymmetry or a secondary emission component. Stacking all spaxels for which nebular emission line flux was detected, we obtain the composite spectrum of BX442 shown in Figure S4 (lower panel). Fitting a single Gaussian emission component to this composite spectrum gives a redshift of $z_{H\alpha} = 2.1765 \pm 0.0001$, where the $\Delta z = 0.0001$ ($\sim 10$ km s$^{-1}$) uncertainty corresponds to the statistical uncertainty in the centroid of the single-Gaussian model. If we had instead determined the redshift from the peak wavelength of H$\alpha$ emission in the nucleus we would instead derive $z_{H\alpha} = 2.1775$, suggesting that the systemic redshift uncertainty $\Delta z \approx 0.0005$ (corresponding to $\sim 50$ km s$^{-1}$) is a function of physical morphology. The redshift derived from the composite spectrum is consistent with the previous NIRSPEC-derived redshift of $z_{H\alpha} = 2.1760 \pm 0.0006$ to within 47 km s$^{-1}$.

---

[8] The effective PSF is larger than that achieved for previous Keck/OSIRIS observations of $z \sim 2$ star-forming galaxies[5,3] because of the larger lenslet scale used in the present study.

[9] The spectral resolution of OSIRIS varies from lenslet to lenslet; we consider a lenslet of average spectral resolution.

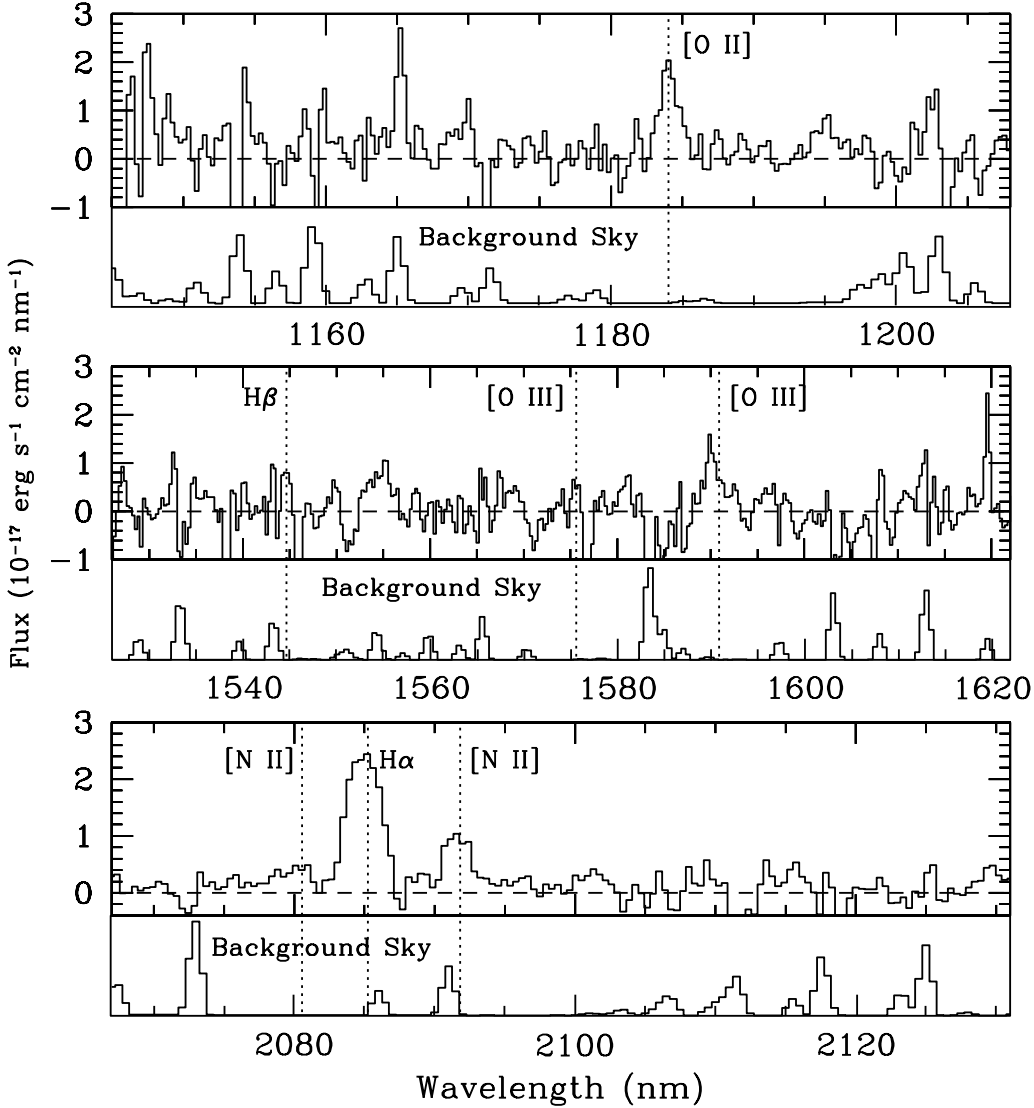

FIG. S3.— **Flux-calibrated $J$, $H$, and $K$-band spectra of BX442 obtained from long-slit Keck/NIRSPEC observations.** Vertical dotted lines denote the fiducial location of major emission line features at the systemic redshift. Insets below each panel show the location and strength of atmospheric OH emission lines (arbitrary normalization) based on a model of the near-IR sky background. Note that the spectral resolution of both the NIRSPEC data and the sky background is $R \sim 1400$, roughly a factor of two poorer than the OSIRIS spectra plotted in Figure S4.

Table S1. **Keck/NIRSPEC Nebular Line Fluxes**

| Line | f[a] | L[b] |
|---|---|---|
| [O II]$\lambda 3726 + 3729$ | $2.7 \pm 0.4$ | $5 \pm 0.7$ |
| H$\beta$ | $< 1.2$ | $< 1.6$ |
| [O III]$\lambda 5007$ | $1.7 \pm 0.4$ | $2.2 \pm 0.5$ |
| H$\alpha$ | $6.0 \pm 0.3$ | $5.6 \pm 0.3$ |
| [N II]$\lambda 6584$ | $1.6 \pm 0.3$ | $1.5 \pm 0.3$ |

[a]Observed flux in units of $10^{-17}$ erg s$^{-1}$ cm$^{-2}$.

[b]Dust-corrected luminosity in units of $10^{42}$ erg s$^{-1}$.

Some recent studies have claimed to see evidence of a broad component underlying the H$\alpha$ emission. We constrain the existence of such a component by stacking together the spectra of the star-forming regions with $\Sigma_{SFR}$ greater than the mean for BX442 (i.e. with $\Sigma_{SFR} > 0.4 M_\odot$ yr$^{-1}$ kpc$^{-2}$, corresponding to the spiral arms), shifting the individual pixel spectra to remove velocity smearing due to large-scale velocity gradients. We show the resulting spectrum in Figure S5, noting that the composite H$\alpha$ emission line is well described by a single symmetric Gaussian component with $\sigma = 87$ km s$^{-1}$. Assuming that a broad underlying component has a FWHM of 500 km s$^{-1}$, we estimate a $3\sigma$ upper limit on the strength of such a component as $f_{broad} < 1.2 \times 10^{-17}$ erg

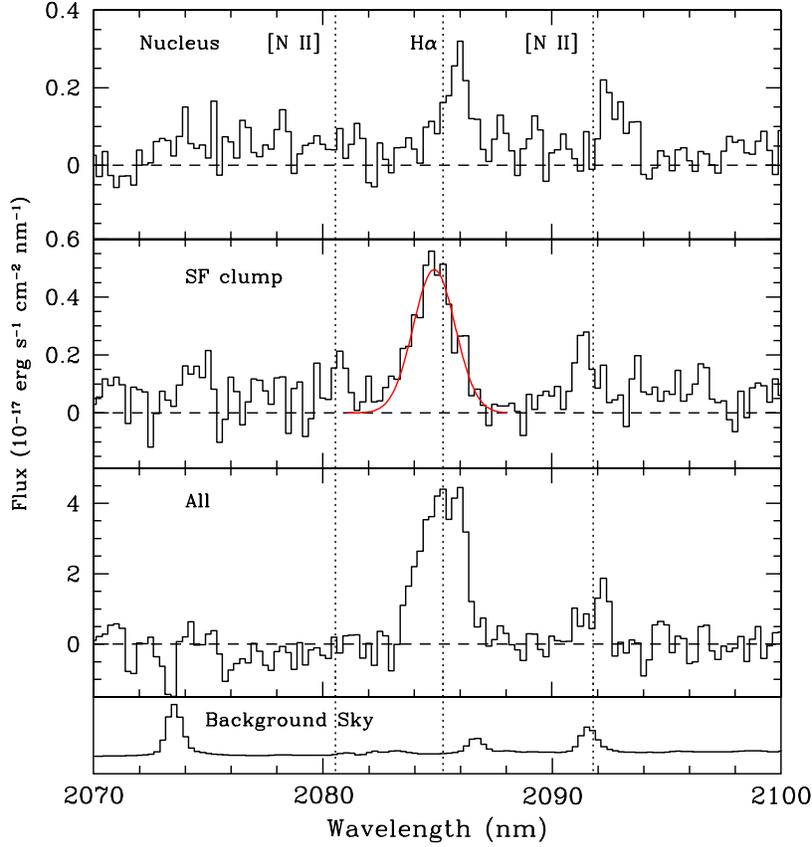

Fig. S4.— **Keck/OSIRIS spectra of Q2343-BX442.** Top panel: spectrum of the central nucleus displaying enhanced [N II] emission. Middle panel: spectrum of the bright star-forming clump located in the northern spiral arm. The red line represents the best-fit single Gaussian emission model. The single Gaussian model is a good fit to the data, with no evidence for a broad wing to the emission line profile potentially indicative of an outflowing wind. Bottom panel: spatially integrated spectrum of the entire galaxy. Note that this is double-peaked and blue asymmetric due to the *physical* morphology of the source; much of the emission comes from the northernmost blue-shifted spiral arm. In all panels the vertical dashed lines denote the fiducial wavelength of [N II] λ6549, Hα, and [N II] λ6585 emission at the systemic redshift of the galaxy; note that this does not line up with the peak wavelengths of emission from the nucleus (top panel) because the systemic redshift is calculated with respect to the total integrated spectrum (bottom panel). The inset at the bottom of the figure represents the background sky spectrum observed by OSIRIS, and indicates the location and strength of atmospheric OH emission lines (arbitrary normalization).

s$^{-1}$ cm$^{-2}$, or $< 20\%$ of the total flux within the summed region. This is in contrast to recent work[21] that found broad, modestly blueshifted components comprising 20-60% of the total Hα flux. However, we note that we are tracing a somewhat different regime of $\Sigma_{\rm SFR}$ as many of the clumps observed previously[21] had $\Sigma_{\rm SFR} = 1-10\, M_\odot$ yr$^{-1}$ kpc$^{-2}$.

Observational uncertainty in both the velocity and velocity dispersion derived in each pixel varies across the galaxy as a complex function of the Hα flux, background sky brightness at the wavelength of peak Hα emission, emission line FWHM, and local velocity gradient. Combined with the correlation of spectra on pixel-pixel scales smaller than the observational point spread function, this makes the observational uncertainty impractical to quantify in every pixel individually. We therefore conducted Monte Carlo tests constructing artificial data cubes mimicking the observed line fluxes, noise statistics, kinematic profile, and spatial correlation introduced by the point spread function smoothing kernel for five locations that were representative of the brightest/faintest, most/least redshifted, and narrowest/broadest Hα emission regions across the galaxy. These tests indicated that uncertainties in centroid velocity and velocity dispersion varied from 13-22 km s$^{-1}$ and 6-16 km s$^{-1}$ respectively in the best and worst cases. For more average pixels the spread in values was smaller (15-20 km s$^{-1}$ and 12-15 km s$^{-1}$ respectively), leading us to adopt source-averaged uncertainties of 17 km s$^{-1}$ in the velocity centroid and 14 km s$^{-1}$ in the velocity dispersion recovered from individual pixels.

Although the Keck II LGSAO system has high relative astrometric precision, the absolute pointing accuracy is insufficient to permit us to automatically align the OSIRIS and *HST*/WFC3 maps of BX442 to better than $\sim 0.2$ arcsecond. Since the central nuclear bulge visible in the WFC3 imaging data has no obvious counterpart in the Hα flux map it is not possible to use this feature to align the maps. Given the similarity of the spiral arm morphology in the Hα and WFC3 imaging, we instead assume that the arms as seen in the *HST* and Keck data are coincident and use these features to align the maps to $\sim 0.1$ arcsec accuracy (Figure S6).[10] This image registration results in the peak [N II] emission region being offset from the F160W nuclear bulge by $\sim 0.1$ arcsec, roughly half the width of the observational PSF. Such spatial offsets of the [N II] peak from the apparent galaxy center have previously been noted in $z \sim 1.5$ star-forming galaxies[12]; if we had instead chosen to register our images based on the [N II] emission peak the spiral arms would be visibly offset from one another.

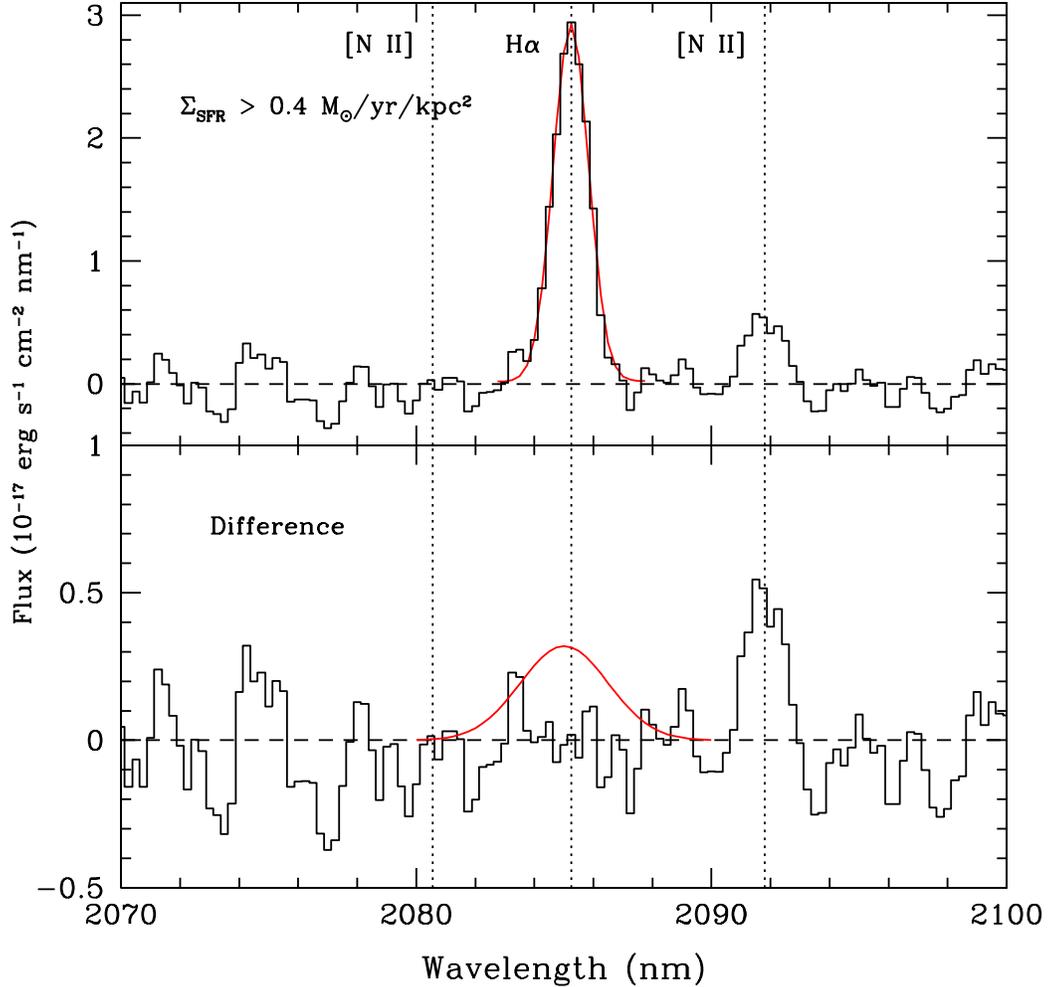

Fig. S5.— **Coadded Keck/OSIRIS spectra of BX442 shifted to the systemic velocity.** Top panel: Coadded spectrum (black histogram) of bright star-forming regions ($\Sigma_{\rm SFR} > 0.4 M_\odot$ yr$^{-1}$ kpc$^{-2}$) within BX442 after shifting the individual pixel spectra to the systemic velocity. The H$\alpha$ emission line in the coadded spectrum is well-described by a single symmetric Gaussian component (red curve). Bottom panel: Residual spectrum after subtraction of the best-fit single Gaussian model for H$\alpha$ emission. Red curve in the bottom panel represents the 3$\sigma$ upper limit imposed on a putative broad component (FWHM $\sim$ 500 km s$^{-1}$) by the noise characteristics of the residual spectrum. In both panels the vertical dashed lines denote the fiducial wavelength of [N II] $\lambda$6549, H$\alpha$, and [N II] $\lambda$6585 emission at the systemic redshift of the galaxy.

### 5. STELLAR POPULATIONS AND GAS CONTENT

We estimate the stellar mass of BX442 by fitting stellar population models to broadband magnitudes obtained from our extensive ground-based $U_n \mathcal{GRJK}_s$,

---

[10] We note, however, that some amount of offset between the H$\alpha$ and F160W arms may be intrinsic to the galaxy, and could in future be used to assess whether the arms of BX442 are leading or trailing.

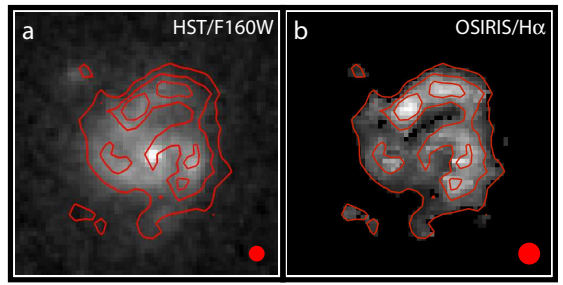

Fig. S6.— **Alignment of HST/F160W and Keck/OSIRIS data.** Panels **a** and **b** show the *HST*/WFC3 and Keck/OSIRIS H$\alpha$ images respectively with red overlaid contours derived from the H$\alpha$ flux map.

*HST*/WFC3 F160W, and *Spitzer* Infrared Array Camera (IRAC) photometry. This SED fitting is described in detail elsewhere[45,47,48]; in brief, we use Charlot & Bruzual (2012, in prep.) population synthesis models, a Chabrier[49] initial mass function, and a constant star formation history. The $H_{160}$ magnitude is corrected for line flux using the [O III] and H$\beta$ fluxes measured from long-slit NIRSPEC spectra, with a factor of two correction for slit losses.[11] As illustrated by Figure S7, BX442 is well fit by a relatively massive ($M_* = 6^{+2}_{-1} \times 10^{10} M_\odot$), old ($1100^{+1000}_{-500}$ Myr) stellar population with a rapid star-formation rate ($SFR_{SED} = 52^{+37}_{-21} M_\odot$ yr$^{-1}$) and moderate visual extinction ($E(B-V) = 0.3 \pm 0.06$).[12] In comparison, for the 134 other galaxies with spectroscopic redshifts $z = 2.0 - 2.5$ in our morphological catalog, we calculate[13] $\langle M_* \rangle = 7 \times 10^9 M_\odot$, $\langle SFR_{SED} \rangle = 20 M_\odot$ yr$^{-1}$, $\langle$Age$\rangle = 340$ Myr, and $\langle E(B-V) \rangle = 0.18$. BX442 is therefore substantially more massive, older, dustier, and is forming stars at a faster rate than the parent UV-selected galaxy population at $z \sim 2$.

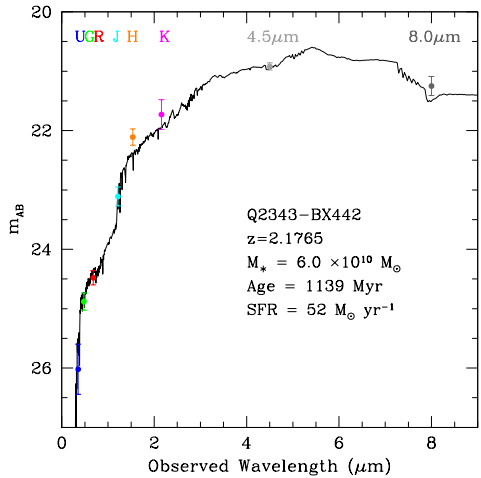

FIG. S7.— **Spectral energy distribution of BX442 based on broadband photometry.** Colored points for $U_nG\mathcal{R}JK_s$ magnitudes represent ground-based optical and near-IR photometry, the orange $H$ magnitude represents $HST$/WFC3 F160W photometry, and the grey points represent $Spitzer$/IRAC observations. The best-fit constant star-formation (CSF) model is overplotted (solid black line). Values given for the stellar mass, population age, and star-formation rate are derived from the best-fit CSF model using a Chabrier[49] IMF.

A secondary estimate of the SFR is provided by the rest-frame mid-IR flux $f_{24} = 162.2 \pm 4.7$ $\mu$Jy measured from $Spitzer$/MIPS 24$\mu$m observations[48], which suggest that BX442 has a total IR flux intermediate between a LIRG and a ULIRG. Converting the $k$-corrected rest-frame 8$\mu$m luminosity $L_8$ to total IR luminosity $L_{IR} = 2.4 \times 10^{12} L_\odot$ using the calibration derived from observations of $\sim 400$ $z \sim 2$ star-forming galaxies[48], the IR star-formation rate relation[51] in combination with a Chabrier[49] IMF indicates that the SFR of BX442 is $SFR_{IR} = 230 M_\odot$ yr$^{-1}$. If we were to adopt a different $L_8$ to $L_{IR}$ calibration that more fully samples galaxies in the ULIRG regime[52] we would instead derive $SFR_{IR} = 85 M_\odot$ yr$^{-1}$. Both estimates of the IR-derived SFR are somewhat higher than SFR$_{SED}$, indicating either that a substantial amount of star formation is occurring in dust-obscured regions or that there is a contribution to the IR luminosity from the faint central AGN whose existence is indicated by spatially resolved [N II]/H$\alpha$ emission line diagnostics (§7).

It is also possible to estimate the current SFR and total gas mass of BX442 from the H$\alpha$ emission by employing the H$\alpha$ star-formation rate relation[51]. Stacking all spaxels in which BX442 is detected into a composite spectrum (Figure S4, lower panel), we estimate its total H$\alpha$ flux to be $f_{H\alpha} = 1.1 \times 10^{-16}$ erg s$^{-1}$ cm$^{-2}$. This estimate is consistent with previous long-slit NIRSPEC observations ($f_{H\alpha} = 6.0 \pm 0.3 \times 10^{-17}$ erg s$^{-1}$ cm$^{-2}$) to within a factor of two correction for slit losses. Adopting a color excess $E(B-V) = 0.3$ and the Calzetti[53] attenuation law, we estimate the dust-corrected H$\alpha$ luminosity of BX442 to be $L_{H\alpha} = 10.2 \times 10^{42}$ erg s$^{-1}$, corresponding to SFR$_{H\alpha} = 45 M_\odot$ yr$^{-1}$.[14] Using the global Schmidt-Kennicutt law to relate the H$\alpha$ surface brightness in each spaxel to the corresponding gas surface density[45,20,54,55], we estimate that the gas mass of BX442 is $M_{gas} = 2 \times 10^{10} M_\odot$. The assumptions implicit in this estimate are significant, and we therefore estimate that $M_{gas}$ is uncertain by at least a factor of two.

A corresponding estimate of the cold gas mass (H$_2$+ He) of BX442 obtained from IRAM Plateau de Bure[56] observations of the CO 3-2 rotational transition suggests[57] a significantly greater gas content $M_{gas} = 1.0 \pm 0.2 \times 10^{11} M_\odot$, comparable to the total $dynamical$ mass interior to 8 kpc. The mismatch between the gas mass estimates may indicate a breakdown in the $X_{CO}$ conversion factor used to estimate the H$_2$ gas mass from the observed CO luminosity.[58] Regardless, BX442 is clearly a gas-rich system, with a gas to total baryonic mass fraction $f_{gas} = \frac{M_{gas}}{M_{gas}+M_*} = 25 - 60\%$.

Combining the H$\alpha$ line flux and F160W continuum maps (smoothed to the spatial resolution of the H$\alpha$ maps) it is possible to investigate the spatial variation in stellar population properties across BX442. Taking the F160W continuum map as a rough proxy for the stellar mass distribution with the galaxy (i.e., apportioning the total stellar mass across the galaxy based on the fraction of the total F160W flux within a given pixel), and converting the H$\alpha$ flux map to a map of spatially resolved SFR[51], the ratio between the two roughly characterizes the stellar population age, or the time required (at the present rate of star formation) to build up the stellar mass present at each location in the galaxy. Figure S8 illustrates that while the average age of stars throughout the galaxy is consistent with estimates obtained via SED-fitting ($1100^{+1000}_{-500}$ Myr), there is a significant age gradient within the galaxy.[15] While the spiral arms are relatively young (varying from 1 Gyr down to 600 Myr in the brightest star-forming knots), the nuclear bulge and interarm regions are significantly older ($\sim 3$ Gyr) and consistent with the age of the Universe at this epoch (3.09 Gyr). All of these ages are significantly longer than the estimated coherence timescale of the spiral features ($\sim 50-100$ Myr), indicating that the tidal forces must have compressed both gas and stars into the arms as the stellar population would not have time to form *in situ* at the present rate of star formation.

---

[11] Since BX442 has an observed magnitude $H_{160} = 22.04$, corresponding to $\sim 2 \times 10^{15}$ erg s$^{-1}$ cm$^{-2}$ within the F160W bandpass, the small correction due to the [O III] + H$\beta$ flux of $\lesssim 5 \times 10^{-17}$ erg s$^{-1}$ cm$^{-2}$ has minimal impact.

[12] We estimate uncertainties on each of these parameters using a series of Monte Carlo simulations in which the observed photometry of BX442 is randomly perturbed according to the photometric uncertainty, and the best-fit SED is recalculated incorporating the star-formation timescale $\tau$ as a free parameter[45,47].

[13] Averages in the stellar mass, SFR, and stellar population age are computed with respect to their base 10 logarithm. Values are slightly below previous estimates in the literature for similar samples[45] because we adopt Charlot & Bruzual (2012, in prep.) instead of Bruzual & Charlot[50] stellar population models.

[14] If the ionized gas is more attenuated than the starlight, $E(B-V)_{star} = 0.44 E(B-V)_{gas}$[53], then we derive SFR$_{H\alpha} = 144 M_\odot$ yr$^{-1}$.

## 6. KINEMATIC DISK-FITTING ALGORITHM

We model the velocity field of BX442 (Figure 2) by constructing a rotating disk model convolved with the observational PSF as follows. We first construct a high-resolution face-on disk model with total diameter of 3 arcseconds on a 0.01 arcsecond grid, assigning each pixel with radius $r$ an H$\alpha$ flux $f(r)$, vertical velocity dispersion $\sigma_z$ perpendicular to the disk, and rotation velocity $v_{\rm rot}(r)$. We adopt a piecewise rotation curve characterized by a solid-body rise from $v = 0$ km s$^{-1}$ at the center to $v = v_{\rm max}$ at a characteristic radius $r_{\rm disk}$, and a flat rotation curve $v = v_{\rm max}$ at $r \geq r_{\rm disk}$. Such rotation curves have been found to be reasonable models for $z \sim 2$ disk galaxies; our analysis is largely unchanged if we instead use a rotation curve of the form $v = \frac{2}{\pi} v_{\rm max} \arctan(r/r_{\rm disk})$[59]. At each position on the grid we construct a mock H$\alpha$ emission spectrum consisting of a single Gaussian profile with centroid and FWHM determined respectively by the relative velocity $v_{\rm los}$ and velocity dispersion $\sin(i)\sigma_z$ projected onto the line-of-sight for a given inclination $i$ (where $i = 0°$ represents a disk viewed face-on), and normalization governed by the relative flux $f$ in the pixel. This data cube is then foreshortened along one axis corresponding to the inclination $i$ (summing spectra where appropriate to mimic the combination in projection of different velocity elements), rotated to a position angle $\phi$, and translated to the desired location $x_{\rm cen}, y_{\rm cen}$ in the field of view. Finally, the data cube is convolved at each spectral slice with a model of the observational PSF, individual spectra are summed into larger effective pixels to match the oversampled spaxel size of the observational data (0.05 arcsec), and the spectrum of each such resampled spaxel convolved with the $R = 3100$ spectral resolution of OSIRIS. Line-of-sight velocity and velocity dispersion maps are recovered from this simulated data cube using automated line-fitting algorithms which perform reliably for such effectively infinite signal-to-noise simulated data.

We calculate the minimum reduced chi square ($\chi^2_r$) of the model minus observed velocity and velocity dispersion maps allowing for possible offsets in the systemic redshift and accounting for the observational uncertainty of the individual measurements, the number of spaxels included in the fit, and the number of degrees of freedom. We exclude non-contiguous features (e.g., the satellite regions of flux to the NE and SE of BX442) from the $\chi^2$ calculation since these are less likely to follow the global velocity field of the disk. We iterate the disk model parameters using a Levenberg-Marquardt $\chi^2$ minimization algorithm implemented in the IDL routine MPFIT2D as detailed below.

Since the observed H$\alpha$ flux map is approximately constant (peaking in regions of higher star formation in the spiral arms, but not displaying a clear radial gradient) we set $f(r)$ equal to a constant. There are therefore 7 free parameters for the model: the peak velocity $v_{\rm max}$, the characteristic disk radius $r_{\rm disk}$, the vertical disk velocity dispersion $\sigma_z$, the inclination $i$ to the line of sight, the position angle $\phi$, and the kinematic center $x_{\rm cen}, y_{\rm cen}$. Given the degeneracy of inclination with rotation velocity, it is not possible to reliably fit for all of these parameters simultaneously. First, we therefore fix reasonable guesses of $i = 30°$ (since the system appears relatively face-on), $\sigma_z = 20$ km s$^{-1}$ (i.e., comparable to local disk galaxies), and iterate the remaining five model parameters. This suffices to strongly constrain the disk position angle $\phi = 168° \pm 1°$ E of N. Defining an ellipse with this position angle, we use the *HST*/WFC3 image of BX442 to estimate the major and minor axes of the ellipse by following the outer isophotal contours of the galaxy (green ellipse in panel **b** of Figure 1). The best visual fit of such an ellipse gives a morphological minor/major axis ratio $b/a = 0.74$; assuming that BX442 has intrinsically circular symmetry and that the flattening is governed entirely by projection, this indicates that the inclination $i = 42°$. This estimate is necessarily crude, predicated as it is on subjective visual analysis and the *a priori* assumption of circular symmetry for the face-on galaxy, and we estimate the corresponding uncertainty in $i$ as $\pm 10°$.

With $i$ and $\phi$ thus fixed, we again iterate our disk fitting algorithm with $\sigma_z = 20$ km s$^{-1}$, and thereby determine the best-fit values for the disk center $x_{\rm cen}, y_{\rm cen}$. Fixing the disk center to these values, we iterate the $\chi^2$ minimization routine again to determine the best-fit values of $r_{\rm disk}$ and $v_{\rm max}$. The optimum value of $\chi^2_r = 2.3$ occurs for values of $r_{\rm disk} = 3.4 \pm 0.1$ kpc and $v_{\rm max} = 234^{+49}_{-29}$ km s$^{-1}$ (the uncertainty in $v_{\rm max}$ is dominated almost entirely by the uncertainty in the inclination angle $i$). The rotation curve of this model is remarkably consistent with that derived for many local disk galaxies, including the Milky Way Galaxy.

Finally, we fix all six parameters except $\sigma_z$, and run the $\chi^2$ minimization routine again, this time calculating the fit between the velocity dispersion maps. The best-fit value is $\sigma_z = 71 \pm 1$ km s$^{-1}$, with a minimum $\chi^2_r = 2.6$. We repeat the steps listed above in order to confirm that our previous best-fit values are unchanged using this revised value of $\sigma_z$. In Figure S9 we plot the 1-dimensional velocity curve for BX442 obtained by resampling the observational data to mimic a long-slit spectro-graph aligned with the kinematic major axis of BX442, along with the corresponding velocity curves of the best-fit piecewise and arctan-based models.

---

[15] Although the derived ages are affected by the uncertainty in image registration between the *HST*/WFC3 and Keck/OSIRIS data, such uncertainty will only be significant on scales significantly smaller than those over which these age gradients are visible.

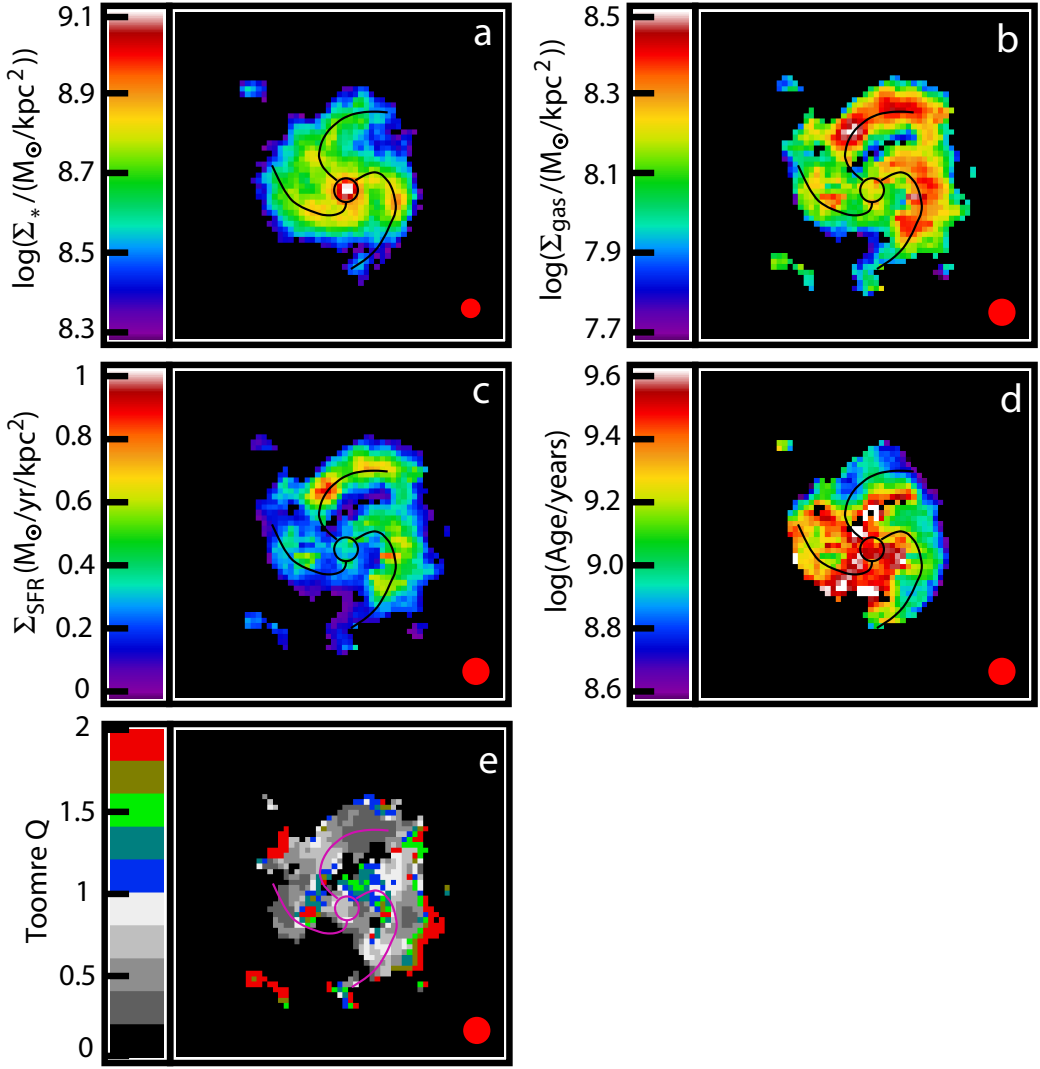

Fig. S8.— **Spatially resolved properties of BX442.** Panels **a**, **b**, **c**: Maps of stellar mass surface density, gas mass surface density, and star-formation rate surface density respectively. Panel **d**: Map of stellar population age constructed from the ratio of the stellar mass and star-formation rate surface density maps. Pixels which have no data in either the F160W or Hα maps are masked out. The age of the universe at the redshift of BX442 is ∼ 9.5 in units of logarithmic years. Panel **e**: Map of the Toomre $Q$ instability parameter throughout BX442. Values $Q < 1$ are unstable to star formation. Red circles in the lower right-hand corner of all panels represent the observational PSF.

Assuming that $v^2 = GM_{\rm dyn}/r$ is indicative of the disk dynamical mass, these parameters suggest that $M_{\rm dyn} = 1.0^{+0.5}_{-0.2} \times 10^{11} M_\odot$ interior to the maximum observed radius $r = 8$ kpc (i.e., the radius at which Hα can no longer be detected in individual spaxels). If we were to instead use a rotation curve of the form $v = \frac{2}{\pi} v_{\rm max} \arctan\left(\frac{r}{r_{\rm disk}}\right)$ our conclusions are largely unchanged; the best-fit rotation curve implies $M_{\rm dyn} = 1.1 \times 10^{11} M_\odot$ at the visible edge of the system. BX442 is therefore intermediate between the local stellar mass Tully-Fisher relation[60] and the 'offset' relation found for $z \sim 2$ galaxies[61], but consistent with both to within observational uncertainty in the deprojected asymptotic velocity.

As indicated by Figure 2, the faint clump of flux to the northeast of the main galaxy is detected in Hα emission and has a relative radial velocity within 100 km s$^{-1}$ of the systemic redshift of BX442. However, the observed clump velocity is discrepant at the ∼ 3σ level with the velocity expected at this position in the best-fit disk model, in contrast to clumps within other $z \sim 2$ disks which have generally been observed to participate in the underlying disk rotation curve[23]. Coupled with its significant angular offset from the rest of the galaxy, we therefore conclude that the faint northeastern clump is likely a small companion in the process of merging with BX442. Although it is difficult to calculate the mass of this companion precisely, estimates of its dynamical mass (based on the observed velocity dispersion and physical size) and stellar mass (based on the percentage of the total F160W continuum flux) indicate that it is probably a few percent

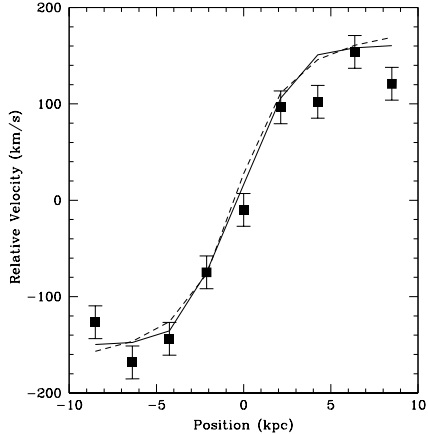

FIG. S9.— **One-dimensional velocity curve of BX442 obtained by resampling the Keck/OSIRIS data.** Simulated pseudo-slit has width 0.25 arcsec and is aligned with the kinematic major axis. Black points represent the observational data with typical uncertainties, spaced in integer steps of the point spread function (0.25 arcsec). The best-fitting disk models obtained using a piecewise solid body + flat and arctan velocity curves are overlaid as solid and dashed black lines respectively.

the mass of BX442.

We note that the best-fit disk center is offset from the apparent center in the HST/WFC3 image by $\sim 2$ kpc (roughly the FWHM of the OSIRIS PSF) spatially and $\sim 100$ km s$^{-1}$ spectrally. This may be due in part to the uncertainty in registering the HST/WFC3 and Keck/OSIRIS maps, but may also indicate a genuine difference between the morphological and dynamical centers. Such offsets are commonly observed in integral-field velocity maps of nearby galaxies[62], particularly for late-type spiral galaxies[63], and are thought to arise because of the high sensitivity of the derived dynamical center to non-circular asymmetries in the rotation curve. The implications of this offset for galaxies in the high-redshift universe have been discussed at some length in the literature[59,61]; in this contribution we permit the dynamical center of BX442 to be a free parameter in order to minimize the effects of non-circular peturbations and recover the most symmetric rotation curve[64].

## 7. SPATIALLY RESOLVED NEBULAR DIAGNOSTIC RATIOS

The composite spectrum of BX442 (defined as the sum of all spaxels in which nebular H$\alpha$ emission is detected) is shown in Figure S4 and exhibits statistically significant [N II] $\lambda 6585$ nebular emission. Defining the quantity $N2 = \log([\text{N II}]/H\alpha)$, we find that the composite spectrum has $N2_{\text{tot}} = -0.70^{+0.10}_{-0.12}$. The uncertainty in $N2$ is estimated by propagating the uncertainties on each of the H$\alpha$ and [N II] line fluxes as determined from artificial spectra in §4. Based on a metallicity calibration relating $N2$ to the oxygen abundance[65]:

$$12 + \log(O/H) = 8.90 + 0.57 \times N2 \quad (2)$$

this ratio implies that the metallicity of BX442 is $12 + \log(O/H) = 8.50^{+0.06}_{-0.07}$, or about 2/3 the solar value $12 + \log(O/H) = 8.66 \pm 0.05$[66]. Given the stellar mass of BX442 ($M_* = 6^{+2}_{-1} \times 10^{10} M_\odot$), the galaxy is therefore consistent with the stellar mass - metallicity relation previously established for $z \sim 2$ star-forming galaxies[67].

However, as indicated by Figure S10 a substantial fraction of the [N II] emission is concentrated in a central clump nearly aligned with the WFC3 nuclear bulge (to within the uncertainty of the image registration). Stacking the spectra of spaxels with measurable [N II] emission in this central clump (see Figure S4, top panel) we find $N2_{\text{nuc}} = -0.22^{+0.13}_{-0.19}$, more consistent with power-law excitation index from a central active nucleus (consistent with the high *Spitzer*/MIPS flux) than from star formation[68]. Such an AGN must necessarily be faint since it is not evident in either the rest-UV spectrum (Figure S2) or the long-wavelength SED (Figure S7). We calculate a dust-corrected H$\alpha$ luminosity for the nuclear region of $L_{H\alpha} \leq 4.2 \times 10^{41}$ erg s$^{-1}$ (where the upper limit indicates that some of the H$\alpha$ flux may be due to nuclear star formation rather than the active nucleus), within the range observed for local Seyfert galaxies[69].

Additional [N II] emission clumps with $\sim 3 - 4\sigma$ detections are located in the inter-arm regions with minimal H$\alpha$ and rest-$g'$ continuum flux (Figure 1, panel **d**). The average value of $N2$ in these clumps is $N2_{\text{ia}} = -0.34^{+0.09}_{-0.10}$. These clumps are therefore either regions of BX442 with comparatively little ongoing star formation and slightly super-solar metallicities, or the [N II] emission may correspond to gas in the warm ionized medium[70] similar to features observed in the inter-arm regions of local spiral galaxies such as M 51[71].

Rejecting all of these spaxels in which significant [N II] emission can be attributed to either AGN heating or to the diffuse ionized medium, we stack the spectra of all spaxels in which [N II] emission is NOT observed individually with confidence $> 3\sigma$ (Figure S10, panel **b**) to estimate the composite spectrum of the star-forming arms of BX442. Although [N II] is not detected in the individual spectra, it is observed in this composite spectrum with $f_{[\text{N II}]} = 1.4 \pm 0.3 \times 10^{-17}$ erg s$^{-1}$ cm$^{-2}$. With a corresponding H$\alpha$ flux of $f_{H\alpha} = 9.3 \pm 0.6 \times 10^{-17}$ erg s$^{-1}$ cm$^{-2}$, this indicates that $N2_{\text{sf}} = -0.82^{+0.08}_{-0.10}$, or $12 + \log(O/H) = 8.43^{+0.05}_{-0.06}$ (i.e., roughly half solar). The metallicity of the actual star-forming regions within BX442 is therefore somewhat less than might be expected from the spatially-averaged stellar mass — metallicity relation[67].

Although we only have spatially resolved information for the H$\alpha$ and [N II] emission lines, by using the spatially integrated Keck/NIRSPEC spectra (Table S1) it is possible to construct the diagnostic emission line ratios $N2_{\text{NIRSPEC}} = -0.57^{+0.08}_{-0.09}$, $\log([\text{O III}]/H\beta) > 0.15$, and $\log([\text{O II}]/[\text{O III}]) = 0.36^{+0.10}_{-0.14}$. The NIRSPEC value for $N2$ is slightly larger than measured for the globally-integrated OSIRIS spectrum, likely because [N II] emission is more strongly concentrated in the nucleus and the NIRSPEC slit therefore covers a greater fraction of the [N II] than the H$\alpha$ emission from BX442. We plot the diagnostic line ratios obtained from both NIRSPEC and OSIRIS in Figure S11. Both the OSIRIS and NIRSPEC globally-averaged diagnostic emission line ratios are con-sistent with a sample of ∼ 100,000 star-forming galaxies drawn from the SDSS[16] and may have an [O III]/[O II] line ratio more like local galaxies than typical $z \sim 2$ star-forming galaxies[73], possibly suggesting a lower ionization parameter. Subtracting the contributions from the faint AGN and warm ionized medium, the regions of peak star formation within BX442 may fall slightly below the local relations, although spatially resolved IFU spectroscopy of [O III] and Hβ emission lines is required to confirm this conjecture. We note that our result concerning the presence of a hitherto-unsuspected faint AGN in BX442 that biases globally integrated nebular diagnostic line ratios is nearly identical to the result found previously for the $z = 1.598$ galaxy HDF-BMZ1299[72]. Indeed, the Hα luminosity of the AGN within BX442 and HDF-BMZ1299 are comparable, estimated at $L_{H\alpha} \leq 4.2 \times 10^{41}$ erg s$^{-1}$ (depending on the contribution of nuclear star formation) and $3.7 \pm 0.5 \times 10^{41}$ erg s$^{-1}$ respectively.

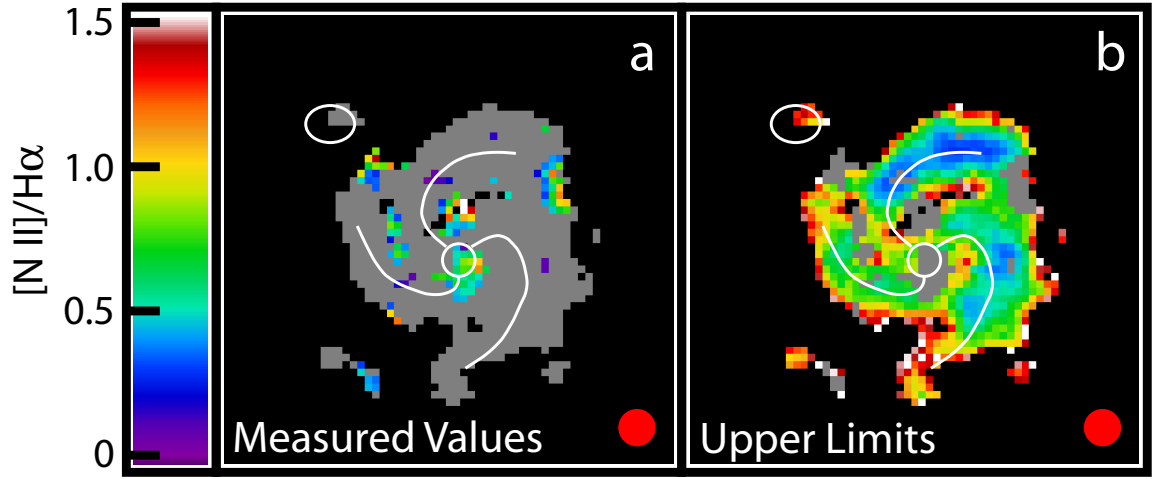

FIG. S10.— **Diagnostic maps of [N II]/Hα line flux ratios.** Panel **a**: Measured [N II]/Hα flux ratios for pixels in which [N II] emission is detected to greater than ∼ 3σ confidence. Panel **b**: Estimated 3σ upper limit on [N II]/Hα flux ratios based on observed Hα emission. Grey regions in panel **a** (**b**) represent pixels for which [N II] emission is not (is) detected to 3σ confidence. Panel **b** indicates that while meaningful constraints on [N II]/Hα cannot be made in the outer edges of BX442, confident limits on [N II]/Hα < 0.5 can be made for the bright star-forming N and W arms. Red circles in the lower right-hand corner of both panels represent the observational PSF.

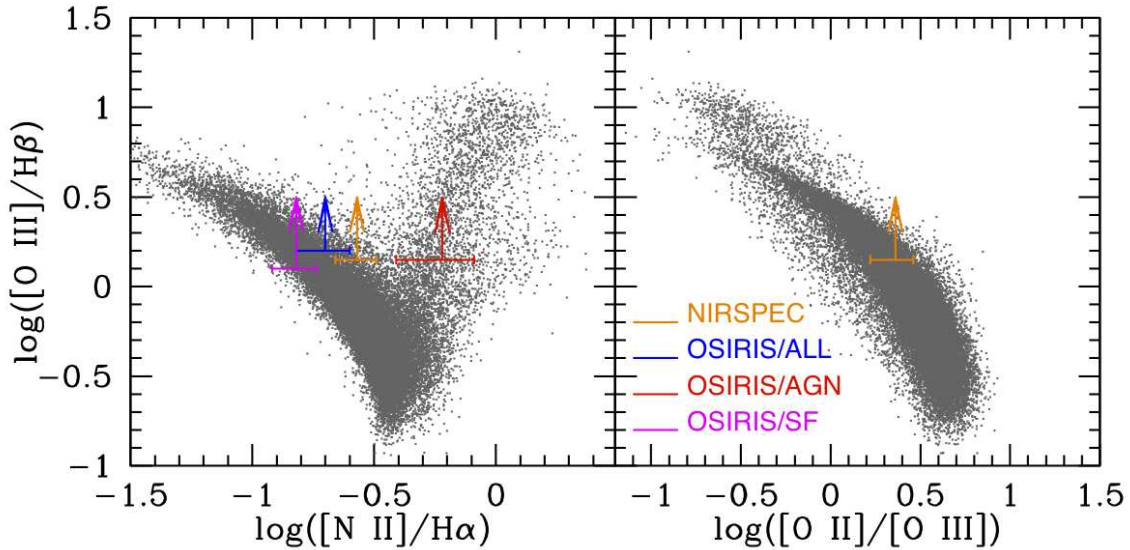

FIG. S11.— **Diagnostic emission line ratio diagram for BX442.** Small grey points represent local galaxies drawn from the SDSS (for clarity, only a randomly selected third of the ∼100,000 galaxies in the SDSS sample are shown). $N2 =\log([\text{N II}]/H\alpha)$ values are given for NIRSPEC long-slit spectra (orange symbols), OSIRIS spatially integrated spectrum (blue), OSIRIS AGN region (red), and OSIRIS star-forming arms (magenta). $\log([\text{O III}]/H\beta)$ values are estimated from long-slit NIRSPEC observations (points are slightly offset for clarity). OSIRIS and NIRSPEC integrated measurements agree within observational uncertainty, but there is significant variation within BX442, as illustrated by red and magenta points.

[16] We use the DR7 catalog available at http://www.mpa-garching.mpg.de/SDSS/, and follow previous studies[72] in selecting galaxies in the redshift range $0.005 \leq z \leq 0.25$ for which [O II], Hβ, [O III], Hα, and [N II] have all been detected with $S/N > 10$.

Given the rapid rate of star formation, we might also expect BX442 to be $\alpha$-element enhanced. As noted previously in the literature[2,74,75], we therefore remark the similarity between BX442 and local thick disks, which tend to be alpha-rich, have slightly subsolar metallicity, and be composed of stars with a vertical velocity dispersion $\sigma_z \sim 65$ km s$^{-1}$ that formed roughly 10 Gyr before the present day[76].

## 8. DYNAMICAL STABILITY

We consider the dynamical stability of BX442 by constructing a map of the Toomre Q parameter[77,78]

$$Q = \frac{\sigma \kappa}{\pi G \Sigma} \quad (3)$$

where $\sigma$ is the velocity dispersion of the ionized gas, $\kappa$ the epicyclic frequency, and $\Sigma$ the local mass surface density. $\kappa$ in turn is given by[78]

$$\kappa^2 = -4B\Omega = -4B(A - B) \quad (4)$$

where $A$ and $B$ are the Oort constants[78]

$$A(R) = \frac{1}{2}\left(\frac{v_c}{R} - \frac{dv_c}{dR}\right) \quad (5)$$

$$B(R) = -\frac{1}{2}\left(\frac{v_c}{R} + \frac{dv_c}{dR}\right) \quad (6)$$

Using our model rotation curve it is possible to calculate $A$, $B$, and therefore $\kappa$ at all points in BX442. We estimate $\Sigma$ as the sum of the stellar and gas mass density at each point, assuming that the total stellar ($M_* = 6^{+2}_{-1} \times 10^{10} M_\odot$) and gas ($M_{\rm gas} = 2^{+2}_{-1} \times 10^{10} M_\odot$) masses estimated for BX442 are distributed according to the relative fluxes in the F160W continuum and H$\alpha$ line flux maps respectively. Finally, we assume that the velocity dispersion $\sigma$ of both stars and gas is given by the characteristic vertical velocity dispersion $\sigma_z$ obtained via our disk-fitting model. The validity of this last assumption is uncertain, both because $\sigma_z$ reflects the velocity dispersion of the ionized gas rather than that of the cold gas or stars, and because the nature and orientation of the velocity dispersion with respect to the disk is unknown[79]. However, we note that if we were to instead use a disk model that adopts an isotropic velocity dispersion and choose to use the radial component $\sigma_r$ in Equation 3 instead, our results for $Q$ would be largely unchanged. Given the various assumptions and uncertainties involved in the calculation of mass surface density, epicyclic frequency, and velocity dispersion, we caution that precise values of $Q$ are probably uncertain at the factor $\sim 2$ level.

We show the spatially resolved map of $Q$ in BX442 in Figure S8. Despite the high velocity dispersion of the system, the mass surface density is sufficiently high that the disk is unstable to fragmentation and star formation ($Q < 1$) in all of the major star-forming regions along the spiral arms. In contrast, there is a region of marginal stability near the core, plausibly consistent (to within centering uncertainties) with the presence of a central nuclear bulge with minimal ongoing star formation. The pixels with $Q \gtrsim 2$ on the outer edges of BX442 represent regions for which BX442 is poorly detected in both F160W continuum flux and H$\alpha$ emission, leading to correspondingly low estimated mass surface densities.

## 9. COMPARISON TO GALAXY FORMATION MODELS

In order to better understand the physical structure of BX442 we compare our results to numerical simulations generated using the *N*-Body Smoothed Particle Hydrodynamic (SPH) code, GASOLINE[29]. GASOLINE includes a redshift-dependent UV background, metal line cooling[80], and non-equilibrium molecular hydrogen abundances. Star formation occurs probabilistically in gas particles as a function of the free fall time and the fractional abundance of molecular hydrogen[81]. Feedback energy from Type Ia and Type II supernovae is distributed to nearby gas particles using a blastwave model[81] that produces galactic winds and regulates the star-formation rate[82]. Similar GASOLINE simulations have reproduced the Tully-Fisher Relation[83], the rotation curves of dwarf galaxies[84], the internal structure of disk galaxies[85–88], the column density distributions and stellar mass-metallicity relation of damped Lyman-$\alpha$ systems at $z \sim 3$[89], and the observed stellar mass-metallicity relationship at $z \sim 0$[90].

The model galaxy is part of a cosmological simulation created using "zoomed-in" initial conditions for which the initial power spectrum of the density field was created using CMBFAST and the simulation was evolved assuming a concordance, flat, $\Lambda$-dominated cosmology with WMAP3[91] values. The dark matter, gas, and star particles have masses of 37,500, 20,000, and 8,000 $M_\odot$, respectively, the force spline softening is 0.17 kpc and the gas smoothing lengths are limited to 0.1 times the softening. At a redshift of zero, these initial conditions result in an approximately Milky Way-mass spiral galaxy.

We used the polychromatic ray-tracing photoionization and radiative transfer program SUNRISE[92,93] to create simulated observations of the galaxy in H$\alpha$ and redshifted F160W emission. As illustrated in Figure S12, the model galaxy at a redshift of $z = 1.86$ exhibits grand design spiral structure similar to that of BX442. At this stage, the simulated galaxy has a stellar mass $M_* = 1.3 \times 10^{10} M_\odot$, gas mass $M_{\rm gas} = 6 \times 10^9 M_\odot$, total luminous disk radius $R = 2.5$ kpc, and a vertical velocity dispersion $\sigma_z = 55$ km s$^{-1}$. The galaxy is actively starbursting, with a total SFR of 12 $M_\odot$ yr$^{-1}$ and peak $\Sigma_{\rm SFR} = 1.4$ $M_\odot$ yr$^{-1}$ kpc$^{-2}$. These characteristics are consistent with the SFR distribution and stellar mass - radius relation for $z \sim 2$ star-forming galaxies[10]. The simulated galaxy also has a nearby companion with stellar mass 2% that of the primary that is 2.9 kpc distant in projection.

This galaxy was selected from a small set of extremely high-resolution cosmological simulations based on its redshift and the presence of an appropriate companion. At this resolution, simulations of massive galaxies such as BX442 are currently prohibitively time intensive, and therefore there are notable differences between BX442 and the simulated galaxy: BX442 is larger ($R \sim 8$ kpc vs 2.5 kpc), more massive ($M_* = 6 \times 10^{10} M_\odot$ vs $1.3 \times 10^{10} M_\odot$), and is forming stars more rapidly (SFR $\sim 50$ $M_\odot$ yr$^{-1}$ vs 12 $M_\odot$ yr$^{-1}$). However, these differences primarily reflect the overall mass normalization of the galaxies; the properties most relevant with regard to dynamical processes such as the excitation of spiral structure are extremely similar. Both galaxies have similar gas fractions ($f_{\rm gas} = 25\%$ vs 32%), merger mass ratios ($\sim 2\%$), impact parameters for the merger relative to the size of the disk ($r_{\rm merg}/R = 1.4$ vs 1.2), and Toomre $Q$ parameters throughout the disk ($Q \sim 0.5 - 1$). Likewise, we expect such dynamical processes as spiral arm formation to be largely independent of the detailed simulation method adopted by GASOLINE (for instance, the use of a supernova blastwave model for stellar feedback in contrast to other models that separately account for momentum feedback from star formation[94,95]), provided the resulting galaxy has a well-formed disk and similar gas fraction and star formation rate to the observed galaxy. Indeed, we note that the generation of spiral structure is robust over the variety of ISM models (e.g., different H$_2$ and/or AGN physics) that we explored.

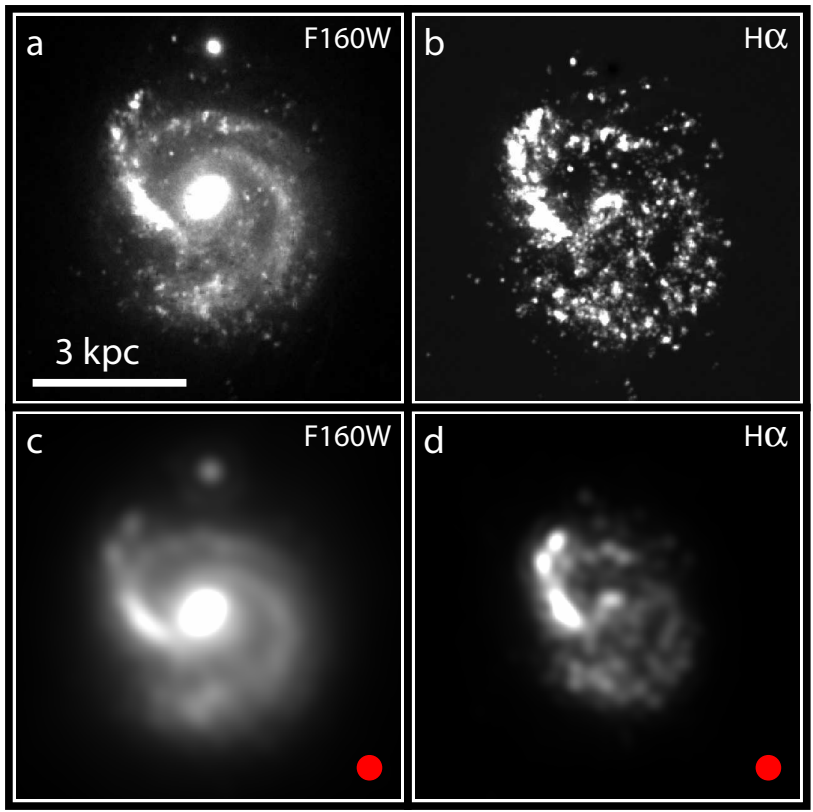

FIG. S12.— **Images of a model $z \sim 2$ spiral disk galaxy produced by a numerical hydrodynamic simulation.** Panels **a** and **b** show simulated morphologies as they would be observed in the F160W broadband filter and H$\alpha$ emission respectively. Panels **c** and **d** are similar to panels **a** and **b**, but illustrate the appearance of the simulated galaxy when smoothed by a mock point spread function (red circles) imitating the angular resolution achieved by our *HST*/WFC3 observations of BX442. The width of the point spread function is defined such that the simulated galaxy has the same number of independent resolution elements across its diameter as BX442. Similarly to observations of BX442 (Figure 1), the model has multiple arms of varying brightness, $\sim 1$ mag arcsec$^{-2}$ arm/interarm brightness contrast in the smoothed continuum image, a continuum-bright nucleus that is largely absent in H$\alpha$ emission, and a faint companion just beyond the disk (2.9 kpc from the simulated galaxy in projection).

By evolving the simulated galaxy forward and backward in time, we observe that the grand design spiral pattern is a transient phenomenon directly related to the orbital phase of the merging companion. Although $Q$ is sufficiently low that the disk spontaneously forms spiral structure in isolation, the spontaneously-formed spiral is flocculent with a low arm/interarm surface brightness contrast that would be indistinguishable at the angular resolution and limiting depth of the *HST*/WFC3 observations of BX442. High contrast two-or three-armed grand design spiral patterns similar to those of BX442 that would be observable with *HST* only appear in the simulation when the merging companion is nearby and close to the plane of the disk; the lifetime of the grand design pattern varies as a function of the orbital parameters of the companion, but is generally less than half a rotation period (i.e., $\leq 100$ Myr, or $\Delta z \leq 0.08$ for BX442).